\documentclass[a4paper,12pt]{article}
\usepackage{amsfonts}
\usepackage{amssymb}
\usepackage{amsmath}
\usepackage{indentfirst}
\usepackage{dsfont}
\usepackage[latin1]{inputenc}
\def\be{\begin{equation}}
\def\ee{\end{equation}}
\def\bea{\begin{eqnarray}}
\def\eea{\end{eqnarray}}

\title{Uniformly accelerated detector in the $\kappa$-deformed Dirac vacuum}
\author{E. Harikumar\footnote{Email: harisp@uohyd.ernet.in}  and Ravikant Verma\footnote{Email:ravikant.uohyd@gmail.com}~\\
{\it School of Physics, University of Hyderabad},\\{\it Central University P O, Hyderabad-500046, India}}

\begin{document}
\maketitle
\begin{abstract}

In this paper, we investigate how a uniformly accelerated detector responds to vacuum state of a Dirac field in the $\kappa$-Minkowski space-time. Starting from $\kappa$-deformed Dirac theory, which is invariant under $\kappa$-Poincare algebra, we derive $\kappa$-deformed Wightmann function for Dirac field, which is valid up to first order in the deformation parameter $a$. Using this, we calculate the response function of the uniformaly accelerated detector, which is coupled to massless Dirac field in $\kappa$-spacetime. From this, we obtain the modification to Unruh effect for the $\kappa$-deformed Dirac field, valid up to first order in the deformation parameter.
\end{abstract}
\section{Introduction}
The non-commutative spacetime was first introduced way back in 1947 by Snyder\cite{sny} in an attempt to harness the UV divergences in the quantum field theories, as suggested by Heisenberg. This idea was resurrected in the last decade due to various developments in non-commutative geometry, string theory and quantum gravity. Combining the well known notions of quantum mechanics and general relativity, it was argued that the structure of the spacetime would be drastically different at extremely short distances of the order of Planck length\cite{dop,dop1}. Further, it was pointed out that the non-commutative geometry provides a possible paradigm to model the microstructure of the spacetime. It was also shown that the usual notions of symmetry and symmetry algebra would be replaced by a Hopf algebra, when the underlying spacetime becomes non-commutative one\cite{chaichian}. The construction of gravity theory on non-commutative spacetime, using the Hopf algebra approach, was studied in \cite{wess,wessa1} and 
many authors have addressed noncommutative gravity and related issues using different approaches\cite{ncgr1,
ncgr1a,ncgr1b,ncgr1c,ncgr1d,ncgr1e,ncgr1f,ncgr1g,ncgr1h,
ncgr2,ncgr2a,ncgr2b,ncgr2c,ncgr3,ncgr3a,ncgr3b,ncgr3c,ncgr3d,ncgr3e,ncgr3f,vor}. 

In recent times, one example of a Lie algebraic type non-commutative spacetime that has attracted wide attention is the $\kappa$-deformed spacetime\cite{l1,l1a,l1b}. This spacetime is defined by the coordinates obeying the commutation relations
\be
[{\hat x}_i, {\hat x}_j]=0,~~[{\hat x}_0,{\hat x}_i]=ia{\hat x}_i,~~a=\frac{1}{k}.\label{com}
\ee

Different field theory models on the $\kappa$-spacetime have been constructed and studied in last couple of years \cite{l1,l1a,l1b,l2,l2a,l2c,l2d,l3,l3a,l3b,dim,dim1a,sm,sm1a,sm1b,sm1c,sm1d,
sm1,sma1,young,young1,young2,us,usa,usb,usc}. 
In \cite{pos,akk,nsri,ab}, the effect of the $\kappa$-deformation in the low energy limits were investigated in different scenarios. Since the $\kappa$-spacetime and $\kappa$-Poincare algebra are known to be related to the doubly special relativity as well as to some models of quantum gravity\cite{jkg}, it is of interest to see how $\kappa$-deformation will affect the motion of particles. The modification of geodesic equation in $\kappa$-spacetime was obtained and analysed in \cite{taj}. It is also of interest to see how the uniformally accelerated observer will respond to $\kappa$-deformation of the spacetime. This question was taken up in \cite{kim,akk2}, where Unruh effect \cite{unruh,unr1,unr1a,unr2,unr2a,unr2b,diracunr,diracunr1,unr3,unr3a} in the $\kappa$-spacetime was studied by analysing the response of the detector coupled to $\kappa$-deformed scalar field. Unruh effect is the fact that a uniformly accelerating detector, coupled to a quantised field which is in its vacuum, will be excited to a 
temperature $T=(2\pi k_B)^{-1}\alpha$, where $\alpha$ is the uniform acceleration of the detector. This is due to that fact that the vacuum of the field theory in the Minkowski spacetime is invariant only under a transformation from one inertial frame to another and not under the transformation from a frame to another which is accelerating\cite{unr2}. Since in the $\kappa$-spacetime, the symmetry algebra is modified from the Poincare algebra, it will be of interest to see whether the Dewitt-Unruh detector will still observe the vacuum as a bath of black body radiation as in the commutative spacetime or not.

It is now well established that the symmetry of the non-commutative space-time and that of the field theory models constructed on such space-time are drastically different from their comutataive analogues\cite{chaichian,wess}. Also field theory models constructed on non-commutative space-time do have many features that are not shared by their commutative counterparts. Since non-commutative space-time and in particular, the $\kappa$-space-time, is expected to be relevant for microscopic theory of gravity, it is of interest to see how well establised results like Unruh effect is modified by the $\kappa$-deformation of the space-time. 

In \cite{akk2}, modification to Unruh effect due to $\kappa$-deformation was analysed for the case of 
scalar field coupled to a detector in the $\kappa$-deformed space-time. This $\kappa$-deformed Klein-Gordon theory was constructed\cite{sm, sm1} in terms of  fields living in the commutative space-time itself and hence the well established calculational tools for field theory could be used in studying the Unruh effect in $\kappa$-spacetime. This $\kappa$-deformed Klein-Gordon theory was constructed as a theory that is invariant under a particular realisation \cite{sm} of the $\kappa$-Poincare algebra, which is the symmetry algebra of the $\kappa$-deformed Minkowski spacetime. It was shown that, for the $kappa$-deformed scalar theory, the Unruh effect gets modified and this effect can be seen in the first order in the deformation parameter itself. Further, this modification was shown to change the Bose-Einstein distribution obtained in the commutative space-time. Thus it is natural to ask what happens to the Unrough effect for the fermions in the $\kappa$-space-time. In this paper, we address this issue.

Using this approach taken in \cite{sm}, $\kappa$-deformed Dirac equation, written in terms of fields defined in the commutative space-time, was derived and analysed in \cite{nsri}. In this paper, we start with this $\kappa$-deformed Dirac field interacting with a detector which is uniformly accelerating. By studying the response of this detector, we obtain the modification to Unruh effect due to $\kappa$-deformation. Unruh effect for the Dirac field was extensively studied in the commutative spacetime, with different motivations\cite{diracunr,unr3}.

This paper is organised as follows. In the next section, we give a brief summary of the $\kappa$-deformed Dirac equation obtained in \cite{nsri}. In section 3, we derived the propagator corresponding to this $\kappa$-deformed Dirac equation, valid up to the first order in the deformation parameter, $a$. Using this, we then introduce the Wightman function in the comoving frame of the uniformly accelerating detector.  In section 4, using this Dirac propagator, we analyse how the uniformly accelerating detector perceive the vacuum of the Dirac field in the $\kappa$-deformed spacetime. We conclude in section 5.

\section{$\kappa$-Deformed Dirac Equation}

In this section, we summarise the construction of the $\kappa$-deformed Dirac equation\cite{nsri} which will be used in later section. This $\kappa$-deformed Dirac equation is written in terms of the operators defined in the commutative spacetime itself. This allows us to use the tools of quantum field theory in the commutative space time to analyse the effect of $\kappa$-deformation. This $\kappa$-deformed Dirac equation is constructed to be invariant under a particular realisation of the $\kappa$-Poincare algebra\cite{sm}. In this realisation, the defining relations of the $\kappa$-Poincare algebra is exactly same as that of the usual Poincare algebra, but the explicit form of the generators are modified. This modifications are dependent on the deformation parameter $a$ and in the commutative limit, these generators smoothly reduces to that of the Poincare algebra. Apart from this modification of the generators, $\kappa$ Poincare algebra differs from the usual Poincare algebra by the fact that it is a Hopf 
algebra.

In the approach taken in \cite{sm}, the $\kappa$-spacetime coordinates are defined as perturbative expansion in the deformation parameter $a$ and obtained in terms of the coordinates of the commutative spacetime and their derivatives as,
\be
\hat{x}_\mu = x^{\alpha} \Phi_{\alpha \mu}(\partial).
\ee
In particular we demand the realisation of the coordinates as
\bea
\hat{x}_i&=&x_i \varphi(A).\label{xi}\\
\hat{x}_0&=&x_0 \psi(A) + i a x_i \partial_i \gamma(A) ,\label{x0}
\eea
where $A=-ia\partial_0$. Demanding that these realisations should satisfy the commutation relations in Eqn.(\ref{com}) leads to the condition 
\be
\frac{\varphi^\prime}{\varphi} \psi=\gamma - 1,
\ee
where we have used $\varphi^{\prime}=\frac{d\varphi}{dA}$. Imposing boundary conditions, we find that   $\varphi(0)=1,$ $ \psi(0)=1$ and $\gamma(0)=\varphi'(0) + 1$, and these functions are finite and  positive. The realisation of ${\hat x}$ for a specific choice of $\psi$ will be characterised by $\varphi$.

The generators of the $\kappa$-Poincare algebra, $D_\mu$ and $M_{\mu\nu}$ are given in terms of the above realisations of the coordinates of the $\kappa$-deformed spacetime as
\begin{equation}
D_i=\partial_i \frac{e^{-A}}{\varphi},~~ D_0=\partial_0 \frac{sinhA}{A} + i a \bigtriangledown^2 \frac{e^{-A}}{2 \varphi^2}. \label{dderivative}
\end{equation}
and
\bea
M_{ij}&=&x_i\partial_j-x_j\partial_i\\
M_{0i}&=&x_i\partial_0\varphi \frac{e^{2A}-1}{2A}-x_0\partial_i(\varphi)^{-1} +iax_i\partial_k\partial_k(2\varphi)^{-1}-iax_k \partial_k \partial_i \frac{\gamma}{\varphi}.
\eea
These generators  satisfy 
\begin{equation}
[M_{{\mu}{\nu}} , D_{\lambda}]=\eta_{{\nu}{\lambda}} D_\mu - \eta_{{\mu}{\lambda}} D_\nu , ~~[D_\mu , D_\nu]=0,\label{kp1} 
\end{equation}
\begin{equation}
[M_{{\mu}{\nu}}, M_{{\lambda}{\rho}}]=\eta_{{\mu}{\rho}} M_{{\nu}{\lambda}} + \eta_{{\nu}{\lambda}} M_{{\mu}{\rho}} - \eta_{{\nu}{\rho}} M_{{\mu}{\lambda}} - \eta_{{\mu}{\lambda}} M_{{\nu}{\rho}} ,\label{kp2}
\end{equation}
where we have used $\eta_{{\mu}{\nu}}$ = diag(-1,1,1,1). The (Dirac) derivatives $D_\mu$ defined above, transform as vectors (contrary to the usual derivative operators in the $\kappa$-Minkowski spacetime). It is easy to see that the Casimir of the $\kappa$-Poincare algebra is $D_{\mu}D^{\mu}$ which can be expressed as
\begin{equation}
D_{\mu}D^{\mu}=\square (1 - \frac{a^2}{4} \square) .
\end{equation}
Here the $\square$ operator is given explicitly as
\be
\square=\bigtriangledown^2 \frac{e^{-A}}{ \varphi^2} + 2 \partial_0^2 \frac{(1-coshA)}{A^2},\label{box}
\ee
where $\bigtriangledown^2=\partial_i \partial_i$ and $A=-i a \partial_0$. 
Note that $\partial_i$ and $\partial_0$ are the derivatives corresponding to the commutative space and time coordinates and these do not transform as vectors under $\kappa$-Poincare algebra while the $D_\mu$ transform as a vector. It is clear that the Casimir, $D_{\mu}D^{\mu}$ reduce to the usual relativistic dispersion relation in limit $a\rightarrow0$. $\varphi$ appearing in above equation, characterizes arbitrary realization of the $\kappa$-spacetime coordinates in terms of commutative coordinates and their derivatives.  $\kappa$-deformed Klein-Gordon equation is written in terms of the above  Casimir of the $\kappa$-Poincare algebra as 
\begin{equation}
\square (1 - \frac{a^2}{4} \square) \Phi(x) - m^2 \Phi(x)=0 .\label{kkg}
\end{equation}
Note that the scalar field and the operators appearing in the above $\kappa$-deformed Klein-Gordon equation are defined in the commutative spacetime.  In the momentum space above Eqn.(\ref{kkg}) becomes
\begin{equation}
\frac{4}{a^2} sinh^2(\frac{ap_0}{2}) - p_{i}^{2} \frac{e^{-ap_0}}{\varphi^2(ap_0)} - \frac{a^2}{4} [\frac{4}{a^2} sinh^2(\frac{ap_0}{2}) - p_{i}^{2} \frac{e^{-ap_0}}{\varphi^2(ap_0)}]^2 = m^2 .
\end{equation}
In terms of Dirac derivatives, $\kappa$-Dirac equation is obtained in\cite{nsri} as
\begin{equation}
(i\gamma^0 D_0 + i\gamma^i D_i + m)\Psi=0.\label{kdirac}
\end{equation}
Note that the $\gamma$ matrices appearing above do not get any $\kappa$ dependent corrections and all the $\kappa$ dependence of the above Dirac equation appear through the Dirac derivatives alone. It is easy to see that the square of above equation gives the $\kappa$-deformed Klein-Gordon equation given in Eqn.(\ref{kkg}).  It was shown that this $\kappa$-deformed Dirac equation breaks the charge conjugation invariance whereas both parity and time reversal continue to be symmetries of Eqn,(\ref{kdirac}). The modification of the Hydrogen atom spectrum was also analysed using the above Dirac equation in \cite{nsri}.

In this paper, we study the modification to Unruh effect for the case of $\kappa$-deformed Dirac field. For this,  we investigate the response  of a uniformly accelerating detector coupled to the $\kappa$-deformed Dirac fermions. The modification to Unruh effect is obtained here is valid up to first order in the deformation parameter $a$. With this in mind,  we start with the $\kappa$-Dirac equation valid up to first order in $a$. For  massless fermions, with the choice $\varphi=e^{- \frac{A}{2}}$, we get the $\kappa$-Dirac equation valid up to first order in the deformation parameter $a$ as
\begin{equation}
(i\gamma^{\mu} \partial_{\mu} - \frac{a\gamma^0}{2} \partial_i \partial_i + \frac{a\gamma^i}{2} \partial_0 \partial_i)\Psi=0 \label{kkd}
\end{equation}\\
It is easy to see that in the limit  $a\rightarrow0$, the above equation reduces to the correct commutative Dirac equation. Up to first order in the deformation parameter $a$, different choices of $\varphi(A)$ differ only by a numerical factor and thus the general features of the models will not be dependent on the specific choice of the realisations we use\cite{akk}. From the next section onwards, our analyses are valid only up to first order in $a$.

\section{$\kappa$-Deformed Dirac Propagator}

In this section we obtain the $\kappa$-deformed Wightman function corresponding to the $\kappa$-deformed Dirac equation. The $\kappa$-deformed Dirac equation in Eqn.(\ref{kdirac}) is written in the commutative Minkowski spacetime.

The $\kappa$-deformed Wightman function for Dirac field, $S^{+}(x-x')$ defined in terms of $G^{+}(x-x')$
% the $\kappa$-deformed Wightman function corresponding to the $\kappa$-deformed Klein Gordon theory which has been derived in \cite{akk2}. The propagator for a $\kappa$-deformed Dirac equation is defined 
as
\begin{equation} 
S^{+}(x-x')=\left(  i\gamma^{\mu} \partial_{\mu} - \frac{a\gamma0}{2} \partial_i \partial_i + \frac{a \gamma i}{2}  \partial_0 \partial_i \right)  G^{+}(x-x'),
\end{equation}
where the $\kappa$-deformed Klein-Gordon propagator given by
\begin{eqnarray}
G^{+}(x-x')&=&\frac{1}{(2\pi)^2}  \frac{1}{(\mid x-x'\mid^2 - (t-t')^2)}\nonumber\\ 
&-& \frac{a^2}{4(2\pi)^2} \frac{(\mid x-x'\mid^2 + 3(t-t')^2)}{(\mid x-x'\mid^2 - (t-t')^2)^3} \nonumber\\
&-& \frac{a^2}{(2\pi)^2} \frac{(\mid x-x'\mid^2 + (t-t')^2) (t-t')^2}{(\mid x-x'\mid^2 - (t-t')^2)^4},
\end{eqnarray}
which is valid up to $a^2$ \cite{akk2}. Using this, the $\kappa$-deformed Dirac propagator, $S^{+}(x-x')$, valid up to first order in $a$ is obtained as
\begin{eqnarray}
S^{+}(x-x')&=& \frac{i}{2\pi^{2}} \left[  \gamma^0 \frac{(t-t')}{[(t-t')^2 - |x-x'|^2]^2} - \gamma^1 \frac{(x-x')}{[(t-t')^2 - |x-x'|^2]^2}\right]\nonumber\\
&+& \frac{a\gamma^0}{4\pi^2}\left[   \frac{1}{[(t-t')^2 - |x-x'|^2]^2} +  \frac{4(x-x')^2}{[(t-t')^2 - |x-x'|^2]^3}\right]\nonumber\\
&+& \frac{a\gamma^1}{\pi^2} \left[  \frac{(t-t')(x-x')}{[(t-t')^2 - |x-x'|^2]^3}\right].\label{kdprop}
\end{eqnarray}
It is easy to see in the limit $a\rightarrow 0$, $S^+(x-x^\prime)$ reduces to the correct commutative result.
In \cite{akk2}, it was shown that the $\kappa$-deformed Klein-Gordon propagator $G{+}(x-x')$ have poles which do nothave smooth  commutative limits. But these poles do not contribute to the propagator(up to second order in $a$). Since here the fermionic propagator $S^{+}(x-x')$ is calculated uisng $G^{+}(x-x')$, we see that the $S^{+}(x-x')$ do have smooth commutative limit as $G^{+}(x-x')$.

\section{Detector Response Function}

In this section, we analyse the the response of a uniformly accelerating detector coupled to the $\kappa$-deformed Dirac fermion. The coupling of detector to the fermions described by the 
$\kappa$-deformed Dirac field given in Eqn.(\ref{kdirac}), is described by the interaction Hamiltonian
\begin{equation}
H_{int} = m(\tau)\overline{\tilde{\Psi}}(\tau)\tilde{\Psi}(\tau) ,
\end{equation}
where $m(\tau)$ is the monopole moment operator of the detector. Since for calculating the Unruh effect, we need to couple the detector with Dirac fermions which are in the instantaneously comoving frame (of the detector), we need to apply a boost transformation to the spinor fields. Thus 
the ${\bar{\tilde{\Psi}}}(\tau)$ and $\tilde{\Psi}(\tau)$ appearing in the interaction Hamiltonian above are the 
Fermi-Walker transported spinors. This field is related to $\Psi(x(\tau))$ as 
\begin{equation}
\tilde{\Psi}(\tau) = S_{\tau} \Psi(x(\tau)),
\end{equation}
where the generator of the boost is given as
\begin{equation}
S_{\tau} = e^{\frac{1}{2} \alpha \tau \gamma^0\gamma^1} = cosh(\frac{\alpha \tau}{2}) + \gamma^0\gamma^1sinh(\frac{\alpha \tau}{2}).\label{boost}
\end{equation}
Note here that the boost generator is not modified by the $\kappa$-deformation as the $\gamma$ matrices appearing in the $\kappa$-deformed Dirac equation in Eqns.(\ref{kdirac},\ref{kkd}) are same as that in the commutative spacetime. Thus we see that the corrections to Unruh effect would be independent of the boost transformation.

We consider that the Dirac field is in the Minkowski vacuum $|0_M>$(as in the commutative case) and detector is in its ground state of energy $E_0$.  For the general trajectory taken by the uniformly accelerating detector, it will not remain in its ground state $E_0$, but will make the transition to the excited state of energy $E>E_0$ as the field makes a transition to an excited state $|\psi>$. As in the commutative case, we assume that the time evolution of the detector governed by its Hamiltonian $H_0$ and it is given by
\begin{equation}
m(\tau)=e^{iH_0\tau}m(0) e^{-iH_0\tau}.
\end{equation}
Using the first order perturbation theory, we find the transition probability of the detector exciting from its ground state to the state of energy $E$, to be
\begin{equation}
P =\sum_E |<E|m(0)|E_0>|^{2} R(\tau ,E - E_0),
\end{equation}
where the $R(\tau ,E)$ is the detector response function. This response function is given as
\begin{equation}
R(\tau ,E) = \int^{\tau}_{0} d\tau \int^{\tau}_{0} d\tau' e^{-iE(\tau-\tau')} <0_M|\overline{\tilde{\Psi}}(\tau)\tilde{\Psi}(\tau)\overline{\tilde{\Psi}}(\tau')\tilde{\Psi}(\tau')|0_M> .\label{unruh2}
\end{equation}
The right hand side of the above equation can be re-expressed as product of two-point functions as 
\begin{equation}
<0_M|\overline{\tilde{\Psi}}(\tau)\tilde{\Psi}(\tau)\overline{\tilde{\Psi}}(\tau')\tilde{\Psi}(\tau')|0_M> = Tr[(\tilde{S}^{+}(\tau , \tau'))^2].\label{tp}
\end{equation}
In the above ${\tilde{S}^+(\tau-\tau^\prime})$ is the Fermi-Walker transported propagator  which is calculated in the comoving frame of the detector. This is related to the propagator
$S^+(\tau-\tau^\prime)$ through a boost transformation as
 \begin{equation}
\tilde{S}^{+}(\tau , \tau') = S_\tau S^{+}(\tau, \tau') S_{-\tau'} .\label{fw1}
\end{equation}
This Fermi-Walker transported propagator will be a function of $\tau-\tau^\prime$. Separating the $a$ independent and $a$ dependent parts of $\tilde{S}^{+}(\tau , \tau')$ as
\begin{equation}
\tilde{S}^{+}(\tau , \tau') = \tilde{S}^{+}_{0}(\tau , \tau') + \tilde{S}^{+}_{1}(\tau , \tau'),\label{kkng}
\end{equation}
we find
\begin{equation}
\left[ \tilde{S}^{+}(\tau , \tau') \right]^2 = \left[\tilde{S}^{+}_{0}(\tau , \tau')\right]^2 + \tilde{S}^{+}_{0}(\tau , \tau') \tilde{S}^{+}_{1}(\tau , \tau') + \tilde{S}^{+}_{1}(\tau , \tau') \tilde{S}^{+}_{0}(\tau , \tau'),
\end{equation}
where we have kept terms up to first order in $a$. 

Using the propagator given in Eqn.(\ref{kdprop}), we now calculate the Fermi-Walker transported propagator  valid up to first order in $a$. For this, we first evaluate the propagator in Eqn.(\ref{kdprop}) in terms of the coordinates of the uniformly accelerating detector, i.e, in terms of the Rindler coordinates given by
\begin{equation}
t= \alpha^{-1} sinh\alpha \tau,~~ x=\alpha^{-1} cosh\alpha \tau,~~ y=0=z~.
\end{equation}
Here $\tau$ is the proper time of the detector and $\alpha$ is the constant acceleration of the detector. Using this, we find, explicitly,
\bea
\tilde{S}^{+}(\xi , \eta)^2 &=& - \frac{\alpha^6}{(16\pi^2)^2} \frac{(\gamma^0)^2}{sinh^{6}(\frac{\alpha \xi}{2})} \nonumber\\
&+& \frac{ia \alpha^7}{2(16\pi^2)} \frac{(\gamma^0)^{2} [2cosh(\frac{3\alpha \eta}{2}) - cosh(\frac{\alpha \eta}{2})]}{sinh^{7}(\frac{\alpha \xi}{2})} ,\label{unruh1}
\eea
where $\xi=\tau - \tau'$ and $\eta=\tau + \tau'$. 
Using Eqn.(\ref{unruh1}), we find
\bea
Tr \left[ \tilde{S}^{+}(\xi , \eta) \right]^2 &=& - \frac{4\alpha^6}{(16\pi^2)^2} \frac{1}{sinh^{6}(\frac{\alpha \xi}{2})} \nonumber\\
&+& \frac{2ia \alpha^7}{(16\pi^2)} \frac{\left[2cosh(\frac{3\alpha \eta}{2}) - cosh(\frac{\alpha \eta}{2}) \right]}{sinh^{7}(\frac{\alpha \xi}{2})},
\eea
which is valid up to the first order in deformation parameter $a$. Using above in  Eqns.(\ref{tp}, \ref{unruh2}), we calculate the response function in the limit of large proper time, i.e. in the limit $\tau\rightarrow\infty$.  Thus we get the detector response function as 
\begin{eqnarray}
\lim_{\tau \rightarrow \infty} R(\tau , \overline{E})&=&F(\tau , \overline{E})\nonumber\\
 &=& \frac{1}{60\pi^3} \frac{\tau \overline{E}}{e^{\frac{2\pi \overline{E}}{\alpha}} - 1} \left[ 4\alpha^4 + 5\overline{E}^2 \alpha^2 + \overline{E}^4 \right]\nonumber\\
&+& \frac{a \alpha^3}{32\pi^3} \frac{1}{e^{\frac{2\pi \overline{E}}{\alpha}} - 1}\left[\frac{1}{3} sinh(3\alpha \tau) \left( \frac{104}{45} \overline{E}^2 + \frac{20}{9} \frac{\overline{E}^4}{\alpha^2} -  \frac{4}{45} \frac{\overline{E}^6}{\alpha^4} \right) \right.\nonumber\\
&-& \left. sinh(\alpha \tau) \left( \frac{47}{45} \overline{E}^2 - \frac{4}{9} \frac{\overline{E}^4}{\alpha^2} -  \frac{4}{45} \frac{\overline{E}^6}{\alpha^4}\right) \right]\nonumber\\
&+& \frac{a\alpha^5}{(2\pi)^4} \left[ \left( 2 cosh^{2}(\frac{3\alpha \tau}{2}) - cosh^{2}(\frac{\alpha \tau}{2})\right) {\cal{A}} \right.\nonumber\\
&+& \left. \frac{8}{3} cosh^{2}(\frac{3\alpha \tau}{2}) ~{\cal{B}}  \right]\label{F}.
\end{eqnarray}
In the above $\overline{E}=E-E_0$ . We have also used to definitions
\begin{eqnarray}
{\cal{A}}= \int^\infty_0 \frac{dk}{k}\left[ \frac{\frac{8}{15} |\frac{2\overline{E}}{\alpha} - 2k| + \frac{1}{6} |\frac{2\overline{E}}{\alpha} - 2k|^3 + \frac{1}{120} |\frac{2\overline{E}}{\alpha} - 2k|^5}{e^{|\frac{2\overline{E}}{\alpha} - 2k|\pi} - 1}\right.\nonumber\\
- \left. \frac{\frac{8}{15} |\frac{2\overline{E}}{\alpha} + 2k| + \frac{1}{6} |\frac{2\overline{E}}{\alpha} + 2k|^3 + \frac{1}{120} |\frac{2\overline{E}}{\alpha} + 2k|^5}{e^{|\frac{2\overline{E}}{\alpha} + 2k|\pi} - 1}\right],\label{34}
\end{eqnarray}
and 
\begin{eqnarray}
{\cal{B}} = \int^\infty_0 \frac{dk}{k}\left[ \frac{\frac{2}{3} |\frac{2\overline{E}}{\alpha} + 2k| + \frac{1}{6} |\frac{2\overline{E}}{\alpha} + 2k|^3}{e^{|\frac{2\overline{E}}{\alpha} + 2k|\pi} - 1}\right.\nonumber\\
-\left. \frac{\frac{2}{3} |\frac{2\overline{E}}{\alpha} - 2k| + \frac{1}{6} |\frac{2\overline{E}}{\alpha} - 2k|^3}{e^{|\frac{2\overline{E}}{\alpha} - 2k|\pi} - 1}\right].\label{calb}
\end{eqnarray}
Using this $F(\tau , \overline{E})$, the transition probability (in the large proper time limit) is obtained as
\begin{equation}
P = \sum_E |<E|m(0)|E_0>|^{2} F(\tau , \overline{E}). 
\end{equation}
Using Eqn.(\ref{F}), the rate of transition probability,
\begin{equation}
{\cal{T}}(\tau , \overline{E},a) =\sum_E |<E|m(0)|E_0>|^{2} \frac{dF(\tau , \overline{E})}{d\tau},
\end{equation}
is obtained as
\begin{eqnarray}
{\cal{T}}(\tau , \overline{E},a) &=& \sum_E |<E|m(0)|E_0>|^{2} \left(\frac{1}{60\pi^3} \frac{\overline{E}}{e^{\frac{2\pi \overline{E}}{\alpha}} - 1} \left[ 4\alpha^4 + 5\overline{E}^2 \alpha^2 + \overline{E}^4 \right]\right.\nonumber\\
&+& \left.\frac{a \alpha^4}{32\pi^3} \frac{1}{e^{\frac{2\pi \overline{E}}{\alpha}} - 1} \left[cosh(3\alpha \tau) \left( \frac{104}{45} \overline{E}^2 + \frac{20}{9} \frac{\overline{E}^4}{\alpha^2} -  \frac{4}{45} \frac{\overline{E}^6}{\alpha^4} \right) \right.\right.\nonumber\\
&-&\left.\left. cosh(\alpha \tau) \left( \frac{47}{45} \overline{E}^2 - \frac{4}{9} \frac{\overline{E}^4}{\alpha^2} -  \frac{4}{45} \frac{\overline{E}^6}{\alpha^4}\right) \right]\right.\nonumber\\
&+& \left.\frac{a\alpha^6}{(2\pi)^4} \left[( 3 sinh(3\alpha \tau) - \frac{1}{2} sinh(\alpha \tau)) {\cal{A}} 
+  4 sinh(3\alpha \tau) ~{\cal B} \right]\right),\label{dirac}
\end{eqnarray}
where ${\cal A}$ and ${\cal B}$ are given in Eqns. (\ref{34}, \ref{calb}), respectively. Above equation shows the modification, valid up to the first order in $a$, to the Unruh effect due to the $\kappa$-deformation of the spacetime. We note that the first term in Eqn.(\ref{dirac}) is the same as in the commutative case and remaining terms depend on the $\kappa$-deformation of the spacetime. These terms vanish in the limit $a\to 0$, reproducing the correct commutative limit. The second term in Eqn.(\ref{dirac}) have the same Bose-Einstein distribution as in commutative case but has the overall multiplicative factor which depends on the deformation parameter $a$ and the last two terms show the deviation of Bose-Einstein distribution due to $\kappa$-deformation of spacetime. All the $a$ dependent terms in Eqn.(\ref{dirac}) depends on the detector's proper time $\tau$. This shows that different detectors will detect the different transition rates. This is similar to the result obtained for the scalar theory in 
the $\kappa$-spacetime in \cite{akk2}.

\section{Conclusion}

In this paper, we have analysed the modification of Unruh effect for Dirac field in the $\kappa$-Minkowski spacetime. We started with the $\kappa$-deformed Dirac theory in Eqn.(\ref{kdirac})  which is invariant under the $\kappa$-Poincare algebra. Using the propagator for the $\kappa$-deformed scalar theory obtained in \cite{akk2}, we evaluated the  propagator  for the $\kappa$-deformed fermionic fields obeying Eqn.(\ref{kdirac}). We have then derived the modification to the Unruh effect due to the $\kappa$-deformation by  introducing an interaction between the uniformly accelerating detector and the $\kappa$-deformed Dirac field in the comoving frame of the detector. This modification is
calculated up to first order in the deformation parameter $a$.  We have seen that the  Bose-Einstein distribution is modified by the $\kappa$-deformation. In the limit $a\rightarrow0$, Eqn.(\ref{dirac}) correctly reduces to the commutative result.

In\cite{unr3}, fermionic field satisfying a Dirac equation with random coefficients was coupled to the detector and it was shown that the correction to Unruh effect has a Dirac-Fermi factor. Here we note that the $\kappa$-deformation do not introduce any such correction factors as seen from Eqn.(\ref{dirac}). One place where $\kappa$-deformation could have affected is the Fermi-Walker transport of the fermionic field we used. Here the boost generator would be modified by the noncommutativity as the symmetry transformation is now different from that of the commutatice space-time. But as we have seen in Eqns.(\ref{kdirac},\ref{kkd}), the $\gamma$ matrices are not altered by the $\kappa$-deformation and hence the  generator in Eqn.(\ref{boost}) is independent of $a$.  As in the case of $\kappa$-deformed scalar theory \cite{akk2}, here also we have taken the detector to be same as that in the commutative spacetime. But it is natural to expect that the detector Hamiltonian to have a $a$ dependent correction, and 
this can lead to corrections to the energy eigen values of $H_0$. This may lead to modification of the Bose-Einstein factor appearing in the transition probability. Recently, by comparing the responses of detector for different spacetimes, violation of equivalence principle was studied\cite{doug}. Since the equivalence principle is violated in $\kappa$-spacetime also\cite{akk}, it will be of interest to see how this can be seen by studying the response of the detectors.  In \cite{sm1}, the twisted statistics obeyed by the Klein-Gordon field living on the Kappa space-time was obtained. Since the commutation relations between the creation and annihilation operators has an important role in deriving the Unruh effect using Bogolubov transformation(see \cite{unr1,unr1a} for a discussion), this twisted statistics may lead to interesting modifications to the Unruh effect. The same can happen in the case of Dirac spinor satisfying the kappa-deformed Dirac equation. For analysing this, one has to first obtain the 
twisted statistics obeyed by the Dirac spinors. Studies of these issues are in progress and will be reported separately.

\noindent{\bf Acknowledgment} We thank A. K. Kapoor for usefull discussions and suggestions.


\begin{thebibliography}{99}
\bibitem{sny}H. S. Snyder, Phys. Rev. 71 (1947) 38.
\bibitem{dop}S. Doplicher, K. Fredenhagen and J. E. Roberts, Phys. Lett. {\bf B331}, 39 (1994). 
\bibitem{dop1}S. Doplicher, K. Fredenhagen and J. E. Roberts, Commun. Math. Phys. {\bf 172}, 187 (1995).
\bibitem{chaichian} M. Chaichian, K. Nishijima and A. Tureanu, Phys. Lett. {\bf B568}, 146 (2003).
\bibitem{wess}P. Aschieri, C. Blohmann, M. Dimitrijevic, F. Meyer, P. Schupp and J. Wess,
Class. Quant. Grav. {\bf 22} (2005) 3511.
\bibitem{wessa1}P. Aschieri, M. Dimitrijevic, F. Meyer and J. Wess,Class. Quant. Grav  {\bf 23} (2006) 1883 (2006).
\bibitem{ncgr1} A. H. Chamseddine, G. Felder and J. Frohlich, Commun. Math. Phys. {\bf 155} (1993) 205.
\bibitem{ncgr1a}J. Madore and J. Mourad, Int. J. Mod. Phys. {\bf D3} (1994)221.
\bibitem{ncgr1b} A. Jevicki and S. Ramgoolam, JHEP {\bf 9904} (1999) 032.
\bibitem{ncgr1c}J. W. Moffat, Phys. Lett. {\bf B491} (2000) 345.
\bibitem{ncgr1d}S. Cacciatori, D. Klemm, L. Martucci and D. Zanon, Phys.Lett. {\bf B536} (2002) 101.
\bibitem{ncgr1e} S. Cacciatori, A. H. Chamseddine, D. Klemm, L. Martucci, W. A. Sabra and D. Zanon, Class. Quant. Grav. {\bf 19} (2002) 4029.
\bibitem{ncgr1f} M. A. Cardella and D. Zanon, Class. Quant. Grav. {\bf 20} (2003) L95.
\bibitem{ncgr1g}H. Garcia-Compean, O. Obregon, C. Ramirez and  M. Sabido, Phys.Rev. {\bf D68} (2003) 044015; 
\bibitem{ncgr1h}J. M. Romero and J. D. Vergara, Mod. Phys. Lett. {\bf A18} (2003) 1673.
\bibitem{ncgr2} A. H. Chamseddine, Commun. Math.Phys. {\bf 218}, (2001) 283.
\bibitem{ncgr2a} A. H. Chamseddine, Phys. Lett. {\bf B504} (2001) 33; A. H. Chamseddine, J. Math. Phys. {\bf 44}, (2003) 2534.
\bibitem{ncgr2b} A. H. Chamseddine, Phys. Rev. {\bf D69}, (2004) 024015; H. Nishino and S.Rajpoot, Phys Lett. {\bf B532} (2002) 334.
\bibitem{ncgr2c} B. P. Dolan, K. S. Gupta and A. Stern, Class. Quant. Grav. {\bf 24} (2007) 1647.
\bibitem{ncgr3} B. M. Zupnik, Class. Quant. Grav  {\bf 24} (2007) 15.
\bibitem{ncgr3a} A. P. Balachandran, T. R.G ovindarajan, K. S. Gupta and S. Kurkcuoglu, Class. Quant. Grav  {\bf 23} (2006) 5799.
\bibitem{ncgr3b} A.Kobakhidze,Int. J. Mod. Phys. {\bf A23} (2008) 2541.
\bibitem{ncgr3c} S. Kurkcuoglu and C. Saemann, Class. Quant. Grav. {\bf 24} (2007) 291.
\bibitem{ncgr3d} X. Calmet and A. Kobakhidze, Phys. Rev. {\bf D72} (2005) 045010.
\bibitem{ncgr3e} X. Calmet and A. Kobakhidze, Phys. Rev. {\bf D74} (2006) 047702.
\bibitem{ncgr3f} P. Mukherjee and A. Saha, Phys. Rev. {\bf D74} (2006) 027702.
\bibitem{vor} E. Harikumar and V. O. Rivelles, Class. Quant. Grav.{\bf 23} (2006)7551. 
\bibitem{l1}J. Lukierski, H. Ruegg, A. Nowicki and V. N. Tolstoy, Phys. Lett. {\bf B264} (1991) 331.
\bibitem{l1a} J. Lukierski, A. Nowicki and  H. Ruegg, Phys. Lett. {\bf B293} (1992) 344.
\bibitem{l1b} J. Lukierski and H. Ruegg, Phys. Lett.  {\bf B329} (1994) 189; J. Lukierski, H. Ruegg and W. J. Zakrzewski, Ann. Phys. {\bf 243} (1995) 90.
\bibitem{l2} P. Kosinski, J. Lukierski and P. Maslanka, Phys. Rev. {\bf D62} (2000) 025004.
\bibitem{l2a}M. Dimitrijevic, L. Jonke, L. Moller, E. Tsouchnika, J. Wess and  M. Wohlgenannt, Eur. Phys. J. {\bf C31} (2003) 129.
\bibitem{l2c} M. Dimitrijevic, L. Moller and E. Tsouchnika, J. Phys. {\bf A37} (2004) 9749.
\bibitem{l2d}L. Freidel, J. Kowalski-Glikman and S. Nowak, Phys. Lett. {\bf B648} (2007) 70.
\bibitem{l3}P. Kosinski, J. Lukierski and P. Maslanka, Czech. J. Phys. {\bf 50} (2000) 1283. 
\bibitem{l3a} J. Kowalski-Glikman and S. Nowak, Phys. Lett. {\bf B539} (2002) 126.
\bibitem{l3b} M. Daszkiewicz, K. Imilkowska, J. Kowalski-Glikman and S. Nowak, Int. J. Mod. Phys. {\bf A20} (2005) 4925.
\bibitem{dim} M. Dimitrijevic, F. Meyer, L. Moller and J. Wess, Eur. J. Phys. {\bf C36} (2004) 117.
\bibitem{dim1a}M. Dimitrijevic, L. Jonke and L. Moller, JHEP {\bf 0509} (2005) 068.
\bibitem {sm}S. Meljanac and M. Stojic, Eur. Phys. J. {\bf C47} (2006) 531.
\bibitem{sm1a}S. Kresic-Juric, S. Meljanac and M. Stojic, Eur. Phys. J. {\bf C51} (2007) 229.
\bibitem{sm1b} S. Meljanac and S. Kresic-Juric, J. Phys. {\bf A41} (2008) 235203.
\bibitem{sm1c} S. Meljanac, A. Samsarov, M. Stojic and K.  S. Gupta, Eur. Phys. J. {\bf C53} (2008) 295.
\bibitem{sm1d}S. Meljanac and S. Kresic-Juric, J. Phys. {\bf A42} (2009) 365204.
\bibitem{sm1}T. R. Govindarajan, K. S. Gupta, E. Harikumar, S. Meljanac and D. Meljanac, Phys. Rev. {\bf D77} (2008) 105010.
\bibitem{sma1} T. R. Govindarajan, K. S. Gupta, E. Harikumar, S. Meljanac and D. Meljanac, Phys.Rev. {\bf D80} (2009) 025014.
\bibitem{young}C. A. S. Young and R. Zegers, Nucl. Phys. {\bf B797} (2008) 537.
\bibitem{young1}S. Ghosh and P. Pal, Phys. Lett. {\bf B618} (2005) 243; 
\bibitem{young2} R. Banerjee, B. Chakraborty, S. Ghosh, P. Mukherjee and S. Samanta, Found. Phys. {\bf 39} (2009).
\bibitem{us} E. Harikumar, Europhysics Lett {\bf 90} (2010) 21001; 
\bibitem{usa}E. Harikumar, T. Juric and S. Meljanac, Phys. Rev. {\bf D84} (2011)  085020;
\bibitem{usb}D. Kovacevic, S. Meljanac, A. Pachol and R. Strajn, Phys. Lett. {\bf B 711} (2012) 122.
\bibitem{usc}S. Meljanac, S. Kresic-Juric and R. Strajn, Int. J. Mod. Phys. {\bf A 27} (2012) 1250057.
\bibitem{pos}P. A. Bolokhov and M. Pospelov, Phys. Lett. {\bf B677} (2009) 160.
\bibitem{akk} E. Harikumar and A. K. Kapoor, Mod. Phys. Lett. {\bf A25} (2010) 1991.
\bibitem{nsri} E. Harikumar, M. Sivakumar and N. Srinivas, Mod. Phys. Lett. {\bf A26} (2011) 1103.
\bibitem{ab} A. Borowiec, K. S. Gupta, S. Meljanac and A. Pachol, Europhysics Lett {\bf 92} (2010) 20006.
\bibitem{jkg} J. Kowalski-Glikman, Lect. Notes. Phys. {\bf 669} (2005) 131(also see the reference therin).
\bibitem{taj} E. Harikumar, T. Juric and S. Meljanac, Phys. Rev. {\bf D86} (2012) 045002.
\bibitem{kim} H-C. Kim, J. H. Yee and C. Rim, Phys. Rev. D75 (2007) 045017.
\bibitem{akk2}E. Harikumar, A. K. Kapoor and R. Verma, Phys. Rev. {\bf D86} (2012) 045022.
\bibitem{unruh}W. G. Unruh, Phys. Rev. D14 (1976)870.
\bibitem{unr1} N. D. Birrel and P. C. W. Davies, 'Quantum Fields in Curved Space', Cambridge University Press, 1982.
\bibitem{unr1a}L. C. B. Crispino, A. Higuchi and G. E. A. Matsas, 
Rev. Mod. Phys. {\bf 80} (2008) 787.
\bibitem{unr2} D. W. Sciama, P. Candelas and D. Deutsch, Adv. Phys. {\bf 30} (1981) 327.
\bibitem{unr2a} S. Takagi, Prog. Theor. Phys. Suppl.{\bf 88} (1986) 1.
\bibitem{unr2b} P. Langlois, Ann. Phys. {\bf 321} (2006) 2027.
\bibitem{diracunr} B. R. Iyer and A. Kumar, J. Phys. {\bf A 13} (1980) 469; 
\bibitem{diracunr1} M. Soffel, B. Muller and W. Greiner, Phys. Rev. {\bf D22} (1980) 1935.
\bibitem{unr3}H. Terashima, Phys. Rev. {\bf D60} (1999) 084001.
\bibitem{unr3a} C. H. G Bessa, J. G. Duenas and N. F. Svaiter, Class. Quant. Grav. {\bf 29} (2012) 215011.
\bibitem{doug} D. Singleton and S. Wilburn, Phys. Rev. Lett. {\bf 107} (2011) 081102.
\end{thebibliography}
\end{document}